Article type: Full Article

# Contrast of nuclei in stratified squamous epithelium in optical coherence tomography images at 800 nm

*Si Chen[1], Xinyu Liu[1], Nanshuo Wang[1], Qianshan Ding[2], Xianghong Wang[1], Xin Ge[1], En Bo[1], Xiaojun Yu[3], Honggang Yu[2], Chenjie Xu[4], and Linbo Liu[1,4,*]*

[*]Corresponding Author (liulinbo@ntu.edu.sg)

[1]School of Electrical and Electronic Engineering, Nanyang Technological University, Singapore 639798, Singapore
[2]Department of Gastroenterology, Renmin Hospital of Wuhan University, Wuhan 430060, People's Republic of China
[3]School of Automation, Northwestern Polytechnical University, Xi'an, Shanxi 710072, People's Republic of China
[4]School of Chemical and Biomedical Engineering, Nanyang Technological University, Singapore 637459, Singapore

**Abstract**

Imaging nuclei of keratinocytes in the stratified squamous epithelium has been a subject of intense research since nucleus associated cellular atypia is the key criteria for the screening and diagnosis of epithelial cancers and their precursors. However, keratinocyte nuclei have been reported to be either low scattering or high scattering, so that these inconsistent reports might have led to misinterpretations of optical images, and more importantly, hindered the establishment of optical diagnostic criteria. We disclose that they are generally low scattering in the core using Micro-optical coherence tomography (µOCT) of 1.28-µm axial resolution *in vivo*; those previously reported 'high scattering' or 'bright' signals from nuclei are likely from the nucleocytoplasmic boundary, and the low-scattering nuclear cores were missed possibly due to insufficient axial resolutions (~ 4 µm). It is further demonstrated that the high scattering signals may be associated with flattening of nuclei and cytoplasmic glycogen accumulation, which are valuable cytologic hallmarks of cell maturation.

## 1. Introduction

Optical reflectance microscopy and endoscopy are powerful tools for the screening and diagnosis of epithelial cancers and their precursors, including but not limited to reflectance confocal microscopy (RCM) [1-5], optical coherence tomography (OCT) [6-13] and microscopy (OCM) [14-17], and spectrally encoded confocal endoscopy (SECM) [18, 19]. In particular, if epithelial abnormalities could be efficiently identified and characterized over large mucosal areas by these tools, it would possibly enable timely decision-makings and save substantial pathology costs, thereby allowing further reductions in risks and costs from unnecessary biopsies or resections [20-22]. Towards this goal, it is important to understand the signals from nuclei which carries important diagnostic information on cellular atypia such as increased nucleocytoplasmic ratio, poor cellular differentiation, and irregular cell arrangement.

The scattering characteristics of the nuclei in the stratified squamous epithelium (SSE) have been extensively investigated in the skin, esophagus, and cervix [1, 2, 12, 14-16, 23, 24]. However, excluding those studies that employed exogenous contrast agents such as acidic acid which increases the backscattering of the nuclei [3, 4, 19], there have been inconsistent reports on the scattering characteristics of cell nuclei, resulting in a standing confusion about whether the nuclei are 'high scattering' or 'low scattering' in squamous epithelial cells (keratinocytes). Specifically, both RCM and OCM studies on the human epidermis agreed that nuclei can be identified as dark areas or low scattering structures within the bright cytoplasm *in vivo* [1, 2, 16]. Whereas, when it comes to the human esophagus and cervix, studies using RCM with <1 µm transverse resolution and OCM with ~ 2 µm transverse resolution and ~ 4 µm axial resolution reported that epithelial nuclei were bright or highly scattering relative to the cytoplasm [5, 14, 15, 25].

Several studies on other types of mammalian cells provided evidences that the nuclei are low scattering in live cells [10, 26, 27]. In particular, Tang et al conducted collinear multiphoton fluorescence microscopy-OCM imaging on cultured human glioblastoma cells stained with fluorescence vital dyes; the OCM system had resolutions of 0.5 µm (transverse) and 1.5 µm (axial, in air) so that scattering contributions from individual organelles could be specifically recognized [27]. Their conclusion is enlightening that the subcellular scattering contrast could originate from organelles like mitochondria, plasma membrane and associated actin filaments, and the cytoplasm-nucleus boundary, and there is little contribution to scattering from regions inside the nuclear core [27]. Their study also suggests that the nucleus has relatively homogeneous refractive index in the core which is likely due to the quasi-uniform distribution of chromatin and proteins, and the heterogeneity is mainly on the boundary between the cytoplasm and nucleus [27].

In this study, using Micro-OCT (µOCT) with cellular-level spatial resolutions [28], we clarify that nuclei in both keratinized and nonkeratinized stratified squamous epithelia are generally low scattering in the core except for some pyknotic nuclei at the most superficial layers. More importantly, in the nonkeratinized squamous epithelium such as those in the esophagus and oral mucosa, we could readily detect high scattering signals at the nucleocytoplasmic boundary sandwiching a dark nuclear core which manifest as bright dots within cells in the *en face* view. Our results suggest that those previously reported 'high scattering' or 'bright' nuclear signals in the esophageal and cervical epithelium were most likely from the nucleocytoplasmic boundary and the low scattering nuclear cores were missed possibly due to insufficient axial resolution [5, 14, 15, 25].

## 2. Materials and Methods

### 2.1. Micro-Optical Coherence Tomography System

We have developed a µOCT imaging system, which is composed of an imaging console and a flexible handheld probe suitable for *in vivo* use. The construction of the imaging console is similar to a previously reported setup [29]. In brief, a supercontinuum light source (SC-5,

Yangtze Soton Laser, Wuhan, China) provided a broadband illumination from 650 nm to 980 nm, enabling a measured axial resolution of 1.70 µm in air or ~1.28 µm in tissue (refractive index = 1.33). The output of the light source was coupled into a single-mode fiber (630-HP, Nufern, USA) and split into a sample fiber (630-HP, Nufern, USA) and a reference fiber (630-HP, Nufern, USA) through a free-space beamsplitter (BS008 and PAFA-X-4-B, Thorlabs, USA). The light back reflected from the reference arm and backscattered from the sample arm were combined by the same beamsplitter and directed into a high-speed spectrometer through a single-mode fiber (630-HP, Nufern, USA). The spectrometer was composed of an achromat collimator (f = 30 mm, AC127-030-B, Thorlabs, USA), a transmission grating (960 l/mm, WP-960/900-30, Wasatch Photonics, USA), a camera lens (AF Nikkor 85 mm f/1.8D, Nikon, Japan), and a line scan camera (EV71YEM4CL2014-BA9, E2V, USA). The output of the camera was transferred to the computer via an image acquisition board (KBN-PCE-CL2-F, Bitflow, USA) (**Figure 1**).

We constructed a flexible probe (**Figure 1**) to acquire images *in vivo*. The sample light was collimated by an achromat (f = 15 mm, AC050-015-B-ML, Thorlabs, USA) and focused by a long-working-distance near-infrared objective lens (M Plan Apo NIR 20X, Mitutoyo, Japan) so that the measured lateral resolution was 1.8 µm. The axial field of view, defined as the axial range of 6-dB intensity roll-off, was ~115 µm and the depth of focus was 32.6 µm in air. We positioned a glass window (fused silica with anti-reflection coating, 1 mm thick) ~100 µm before the focal plane to maintain the working distance between the objective lens and the epithelium. The glass window is 2 mm in diameter and mounted on a three-dimensional (3D) printed tapered holder for access to the human labial mucosa *in vivo*. For *ex vivo* studies, this flexible probe was mounted in a microscope frame. The light power on the tissue was 5.5 mW and 1.8 mW for *ex vivo* and *in vivo* studies, respectively.

Each µOCT cross-sectional image contained maximally 2048 axial lines and the data size of each 3D image was 1024 × 1024 × 2048 (x × y × z). The corresponding tissue volume was 1.744 mm × 1.744 mm × 0.88 mm (width × width × depth in water (refractive index = 1.33)). We used an axial line rate of 20k Hz and 60k Hz for *ex vivo* and *in vivo* studies, respectively.

## 2.2. Tissue Preparation and Image Acquisition

In this study, we investigated the scattering characteristics of nuclei in the cultured human keratinocytes, rat epidermis, esophageal and cervical epithelium *ex vivo*, swine epidermis and oral mucosa *ex vivo*, human esophageal epithelium *ex vivo*, and human oral mucosa *in vivo*, respectively.

Primary human epidermal keratinocytes (ATCC PCS-200-010) were incubated with growth medium containing dermal cell basal medium (ATCC PCS-200-030) and keratinocyte growth kit (ATCC PCS-200-040) for 24 hours and then cultured with the growth kit (ATCC PCS-200-040) for another 3-4 days to achieve a density of 75-80%. Thereafter, the cells were seeded in Transwell inserts (Corning 3422) at a density of $5\times10^4$ cells/cm2 and µOCT imaging was performed when a cell density of ~50% was achieved. The Institutional Review Board (IRB) at Nanyang Technological University (NTU) approved the study (IRB 17/12/04).

For rat studies *ex vivo*, 5 Sprague Dawley rats (female, 10-14 weeks) were sacrificed and organs including skin, esophagus and cervix were harvested. µOCT imaging was immediately conducted and all the specimens were fixed with 10% neutral-buffered formalin (Leica Biosystems) for histological analysis. The study was approved by Institutional Animal Care and Use Committee (IACUC) of NTU (ARF-SBS/NIE-A0312).

For swine studies *ex vivo*, we harvested the fresh skin, oral mucosa (floor of mouth) and esophagus from 4 Landrace cross white pigs (female, 12 months) immediately after the cessation of vital signs. Specimens with a size of 1-2 cm × 2 cm were separately stored in serum-free Dulbecco's Modified Eagle's Medium (DMEM), before shipped on ice within 1 hour to the imaging site, where they were preserved in DMEM at 37°C ventilated with 95% oxygen and 5% carbon dioxide. µOCT image acquisition of all specimens was completed within the next 2 hours. The specimens were then fixed for histology (RM2235, Leica Biosystems). The use of swine tissues was approved by the IACUC of National University of Singapore (R14-0797).

For human esophagus studies *ex vivo*, µOCT imaging was performed from the clean margin of 4 esophageal specimens. Following image acquisition, all the specimens were fixed overnight and subject to a standard histology processing protocol. The study was approved by the IRB at Renmin Hospital of Wuhan University (2017K-C053). In addition, we also recruited 2 human subjects and conducted µOCT imaging of Lugol's positive human labial mucosa *in vivo*. The lower labial mucosa was in gentle contact with a food packing film, which coated the fused silica window of the flexible probe (Fig. 1). 3D images were acquired within 17 seconds. 3% Lugol's solution of the lower labial mucosa was applied after µOCT imaging. Written consents were obtained from volunteers before the experiment. The study was approved by the IRB at NTU (IRB-2016-10-015).

## 2.3. Statistical Analysis

The optical intensity was manually measured from 50 randomly selected locations within an area of $2 \times 2$ pixels and $8 \times 8$ pixels (width × height) at each location for nucleus and cytoplasm, respectively. Here we normalize the optical intensity as the ratio between the optical power back-reflected or back-scattered from the intracellular components and that from a perfect reflector (reflectivity: 100%). Since the imaging speed and input power of experiments conducted *in vivo* and *ex vivo* were different, the quantitative reflectance data was normalized to the exposed optical energy to allow direction comparison. All the numerical values are presented as mean ± standard deviation (SD) and $p < 0.05$ was regarded as a statistically significant difference. Independent-samples t-test was used to compare the intensity between the nuclei and cytoplasm of each specimen. One-way analysis of variation (ANOVA) was adopted to compare the intensity of cell nuclei among epithelial tissues and Bonferroni post hoc test was used for further comparisons if statistically significant difference was detected. All statistical analyses were conducted using SPSS software (IBM SPSS Statistics 23.0).

## 3. Results

### 3.1. Cultured Human Keratinocytes

µOCT images of cultured keratinocytes provided cytologic information at a scale comparable to cytology (**Figure 2**). From both cross-sectional (**Figure 2B1**) and *en face* views reconstructed from 3D OCT datasets (**Figure 2B2**), the nuclei demonstrated an intensity of $(0.13 \pm 0.07) \times 10^{-6}$, which was significantly lower than that of the cytoplasm $(1.16 \pm 0.55) \times 10^{-6}$ (t-test; *p < 0.001*). The low-scattering nuclei can be clearly identified including the two nuclei in a keratinocyte undergoing cell division (**Figure 2C&D**). These observations agree well with the previous results in cultured human glioblastoma cells [27].

### 3.2 Keratinized Stratified Squamous Epithelium

We used tissues from albino rats and white swine to exclude the effect of melanin on cytoplasmic scattering [1]. µOCT images acquired from the rat skin, esophagus and cervix and swine skin revealed well-developed keratinized layer (**Figure 3**). Layered structures of mucosal wall and subcellular information of the epithelial cells (keratinocytes) were fully reproduced by µOCT (**Figure 3A1-D1**), matched well with the corresponding histology (**Figure 3A2-D2**). The nuclei of keratinocytes in all types of SSE were generally lower scattering than the cytoplasm (**Figure 3A1-D1,** yellow arrows).

Note that the basal layer are crowded with basal cells which have large nuclei according to histology (large nucleocytoplasmic ratio), so that the basal layer demonstrated relatively low optical intensity in µOCT images (**Figure 3**). As the cells evolve to the middle and upper epithelial layers, they increase in volume particularly due to the accumulation of cytoplasmic inclusions, presenting increased cytoplasmic scattering signals and therefore readily detectable nucleocytoplasmic contrast.

**3.3 Nonkeratinized Stratified Squamous Epithelium**

In the cross-sectional µOCT images from swine floor of mouth (**Figure 4**), we could frequently identify paired high-scattering signals at the nucleocytoplasmic boundary each sandwiching a low-scattering nuclear core in the middle and upper epithelial layers, which corresponded to the Periodic acid-Schiff-Diastase (PAS-D) positive cells with flattened nuclei based on the matched histology (**Figure 4D-E2**, white arrows & arrowheads). In contrast, the high-scattering signals were less often visible in keratinocytes with PAS-D negative cytoplasm and roundish nuclei (**Figure 4A-B2**). The same nuclear scattering characteristics can be found in the cross-sectional images of the human esophageal epithelium *ex vivo* and the Lugol's-positive human labial mucosal epithelium *in vivo* (**Figure 5 A,C,F**, white arrows & arrowheads). These high-scattering signals demonstrated as bright dots in the *en face* views reconstructed from the 3D OCT datasets (**Figure 4F, 5B&D**). When we resliced the µOCT datasets, these paired 'bright' boundaries manifested as twinkling stars against the relatively 'dark' cytoplasm (**Video S1-3**).

It is worth mentioning that cells in the nonkeratinized epithelium is actually keratinized but to a less degree of keratinization, and therefore, their nuclei may undergo pyknosis (condensation of chromatin) or fragmentation and the cytoplasmic organelles could disappear as they migrate to the epithelial surface. This phenomenon of the terminal maturation of keratinized cells could be captured by µOCT in the human labial mucosa *in vivo* where cells presented bright irregular dots within empty ('dark') cytoplasm (**Figure 5E**).

Finally, we measured the scattering intensity of nuclei from some of SSE investigated and normalized the intensity of nuclear cores to a perfect reflector. It can be seen that nuclei were consistently low scattering with respect to their cytoplasm (**Table 1**; Independent-samples t-test). We also compared the optical intensity of nuclei among epithelial tissues, which demonstrated that the nuclei in rat and swine epidermis were statistically higher than those in the rest epithelia, and further Bonferroni post hoc tests showed that, excluding the epidermis, there were no statistically significant differences in nuclear intensity among the rat esophagus and cervix, swine floor of mouth (PAS-negative) and human labial mucosa (One-way ANOVA, $F=47.60$, $p < 0.001$; **Figure 6**).

## 4. Discussion and Conclusion

This study establishes that cell nuclei of SSE can be visualized by their negative contrast against the cytoplasm or the high scattering nucleocytoplasmic boundary. Our study clarifies the discrepancy among the previous observations and provides a consistent and comprehensive interpretation of the optical characteristics of nuclei in SSE. According to the images acquired by µOCT, the axial range of the flattened nuclei are approximately 2.22-2.59 µm, analogous to that in the corresponding histology (~ 2.38 µm), so that the upper and lower scattering signals at the nucleocytoplasmic boundaries may not be resolved in the previous studies due to the limited axial resolution (~ 4 µm) [5, 14, 15, 25], which, consequently, missed the fact that the nuclear cores were low scattering. To validate our understanding, we degrade the axial resolution to 3.7 µm in tissue algorithmically and it is clear that the low-scattering nuclear cores cannot be recognized (**Figure 7**).

While Aguirre. et al mentioned that the nuclei in the human epidermis were not readily visible using OCM [15], several RCM and OCM studies reported the visualization of 'dark' nuclei in human skin, where melanin provides the strong contrast for the cytoplasm [1, 2, 16]. To disclose the native scattering property of nuclei and the nucleocytoplasmic contrast in the intact epidermis, tissues from albino animals or white swine were used in this study. In contrast to images from the pigmented skin that the basal layer demonstrates the brightest intensity due to melanin accumulation [1], our results showed that the basal layer in the non-pigmented SSE presented relatively lower intensity than the middle and upper layers. We found that the nuclei of differentiated cells remained to be low scattering relative to the cytoplasm when melanin was absent, agreed with previous studies in rat bladder and cultured human glioblastoma cells [10, 26, 27]. Note that the scattering signals from the cytoplasm may be too speckly to clearly delineate the negative-intensity areas of nuclei in individual OCT/OCM images, and this speckle issue could be resolved by averaging multiple time-lapsed frames or consecutive *en face* frames [10, 26].

Nuclear flattening and glycogen accumulation are hallmarks of cell maturation in the nonkeratinized SSE such as those in the esophagus and cervix. The high scattering signals were more commonly found in the upper half epithelial layers where an epithelial cell tends to have a more flattened nucleus due to maturation, and this morphological change may be associated with the signals at the nucleocytoplasmic boundary. The current study also suggests that the nucleocytoplasmic high-scattering signals may co-exist with the glycogen accumulation in the cytoplasm (PAS+). Therefore, the paired, high-scattering signals at the nucleocytoplasmic boundary could be used to evaluate cell maturation since the normal maturation processes are disrupted in neoplasia or cancers, including nuclei flattening and glycogen accumulation. It is possible that maturation associated cell events such as glycogen accumulation might have changed the refractive index difference between the cytoplasm and nuclei, as cell nuclei normally have lower refractive index and mass density than cytoplasm [30, 31]. However, the cell physiology is rather complicated and a serial of signal pathways may be activated or repressed responding to the signals for glycogen accumulation, nuclei flattening and other cell maturation related events. Since intracellular glucose could also alter the transcription program of the cells[32, 33], it is possible that the refractive index of the nucleus may be affected owing to the varied level of gene expression. It is the limitation of the current study that the refractive index changes of the cytoplasm and the nucleus during maturation were not studied. Further investigations will focus on analyzing the refractive index of cytoplasm and nucleus of glycogen-rich keratinocytes and exploring the correlation between the OCT signals and the cytologic changes.

This study clarifies that the origin of the high-scattering signals is the nucleocytoplasmic boundary instead of the nuclear core, which lays the basis for the improved understanding that these high scattering signals may indicate cell flattening and can be used as an imaging marker of cell maturation. To demonstrate the benefit of this new imaging marker, validation studies will be conducted on animal models and human samples with epithelial neoplasia against the clinical gold standard. Another unresolved question is why the nuclear cores are low scattering. We have conducted a preliminary study on the ultrastructural characteristics of the keratinocytes in SSEs with transmission electron microscopy. In both keratinized swine epidermis and nonkeratinized SSE from floor of month, the nuclear inclusions were homogenously distributed relative to those in the cytoplasm (**Figure S4**), suggestive of much less refractive index heterogeneity within the nucleus than the cytoplasm.

We conducted a preliminary quantitative analysis of the optical intensity of the nuclear core and the cytoplasm which showed that the former was lower than the latter in each type of SSE. Considering the heterogeneity of cytoplasmic inclusions among epithelial types, we did not analyze the optical intensity of cytoplasm among epithelial types. We only compared the intensity of nuclear cores among tissues which indicated that nuclear optical intensity was mostly similar across types of SSE except for the epidermis. Since we know that the influence of speckles increases with the reflectivity of the sample, we attribute the measured high image intensity from nuclear cores in epidermis to speckles from the cytoplasm or the nucleocytoplasmic boundary, which might have been included in the manually-determined areas of nuclei due to the resolution limitation (**Figure 3A1&D1**). In addition, other imaging artifacts such as attenuation of light by epithelial layers over the measured cell, autocorrelation artifacts from the highly reflective or scattering keratinized layer, lamina propria, or cytoplasm might also have altered the measured values.

In conclusion, the current study establishes the negative contrast of nuclei in the mammalian SSE by use of µOCT in both small and large animals *ex vivo*, and human labial mucosa *in vivo*. We unify most of the previous inconsistent reports on the nuclear scattering characteristics which may contribute to the establishment of a consensus on nuclear scattering properties and new OCT diagnostic criteria based on the subcellular reflectance contrasts. Future works on elucidating the biological background underlying the subcellular scattering contrasts may open up a new avenue towards 'virtual histology' over a large mucosal area *in vivo*.

**Acknowledgements**

This research is supported by the Singapore Ministry of Health's National Medical Research Council under its Cooperative Basic Research Grant (NMRC/CBRG/0036/2013), National Research Foundation Singapore under its Competitive Research Program (NRF-CRP13-2014-05), Ministry of Education Singapore under its Academic Research Fund Tier 1 (2018-T1-001-144), Agency for Science, Technology and Research (A*STAR) under its Industrial Alignment Fund (Pre-positioning) (H17/01/a0/008), and NTU-AIT-MUV program in advanced biomedical imaging (NAM/15005). We thank helpful suggestions from Professor David Sampson on the optical properties of cell nuclei. Si Chen and Xinyu Liu contributed equally to this work.

**Conflict of interest**

The authors declare that there are no conflicts of interest related to this article.

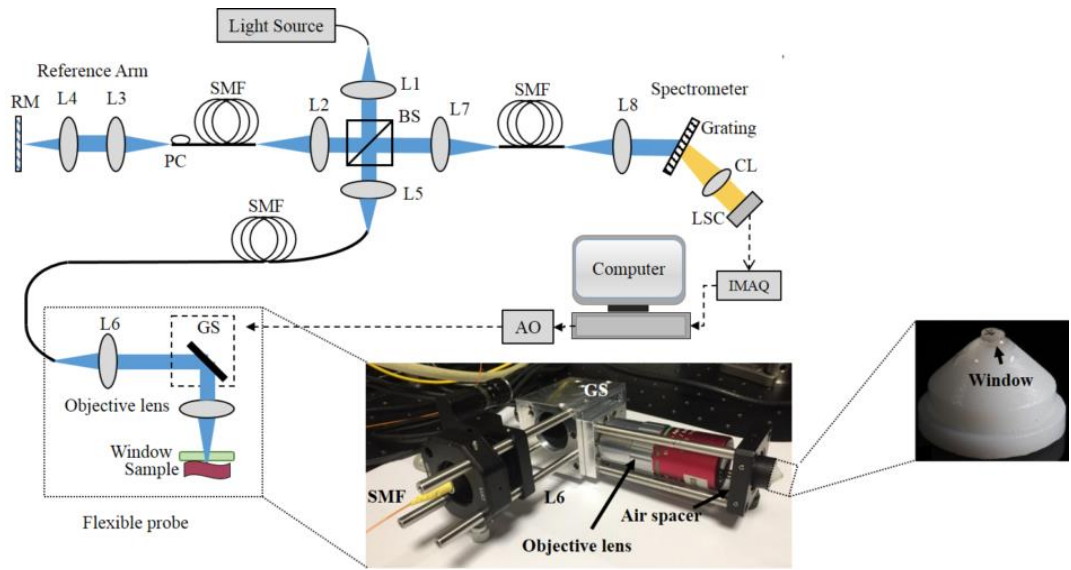

**Figure 1.** A Schematic of µOCT imaging system and handheld probe. L1-L8: optical lenses; BS: beamsplitter; SMF: single mode fibre; GS: galvanometer scanning mirrors; PC: polarization controller; RM: reference mirror; CL: camera lens; IMAQ: image acquisition board; AO: analog outputs.

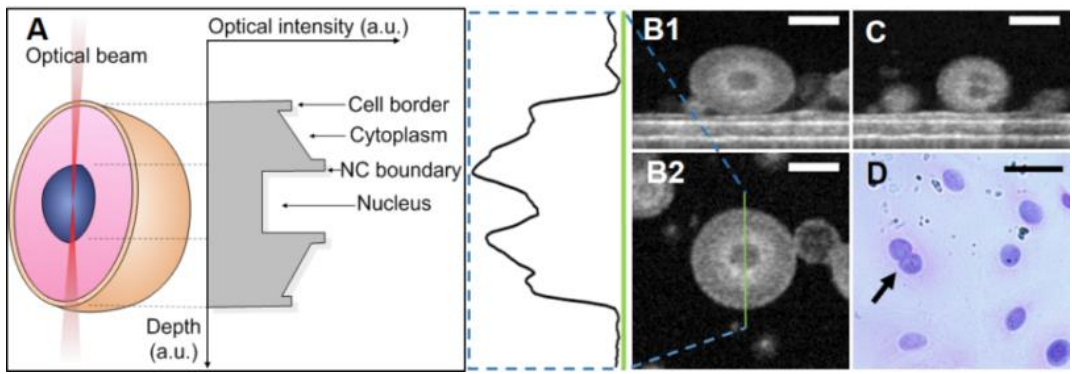

**Figure 2.** Optical intensity distribution in µOCT images of cultured human keratinocytes. (A) A schematic of subcellular optical intensity distribution of cultured keratinocyte. (B) Representative µOCT images of a keratinocyte viewed from the cross-section (B1) and *en face* (B2) view: the optical intensity profile of the keratinocyte (blue dashed box) was derived from real-imaging data in (B2). (C) A keratinocyte undergoing cell division with two nuclei. (D) Representative cytology with a cell that has just completed nuclear division (arrow). NC: nucleocytoplasmic. Scale bars, 30 µm.

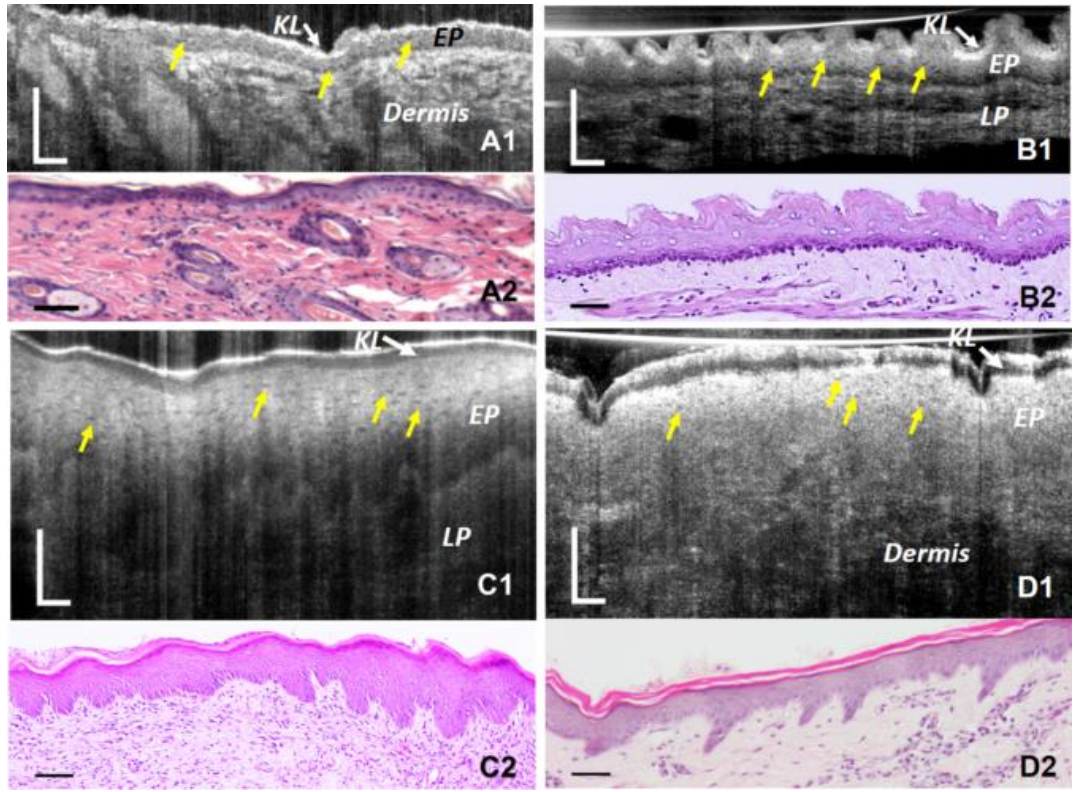

**Figure 3.** μOCT images of the orthokeratinized stratified squamous epithelium in rat and swine acquired *ex vivo* (Images with isotropically sized pixels could be found in **Figure S1**). (A1-D1) Cross-sectional μOCT images of the albino rat skin, esophageal mucosa, cervical mucosa, and skin of white Cross-Landrace swine, respectively; (A2-D2) the corresponding histology with H&E staining. Yellow arrows indicate low-scattering nuclei. The contrast of all images was adjusted with imageJ for the optimal visualization of nucleus. KL: keratinized layer; EP: epithelium; LP: lamina propria. Scale bars, 50 μm.

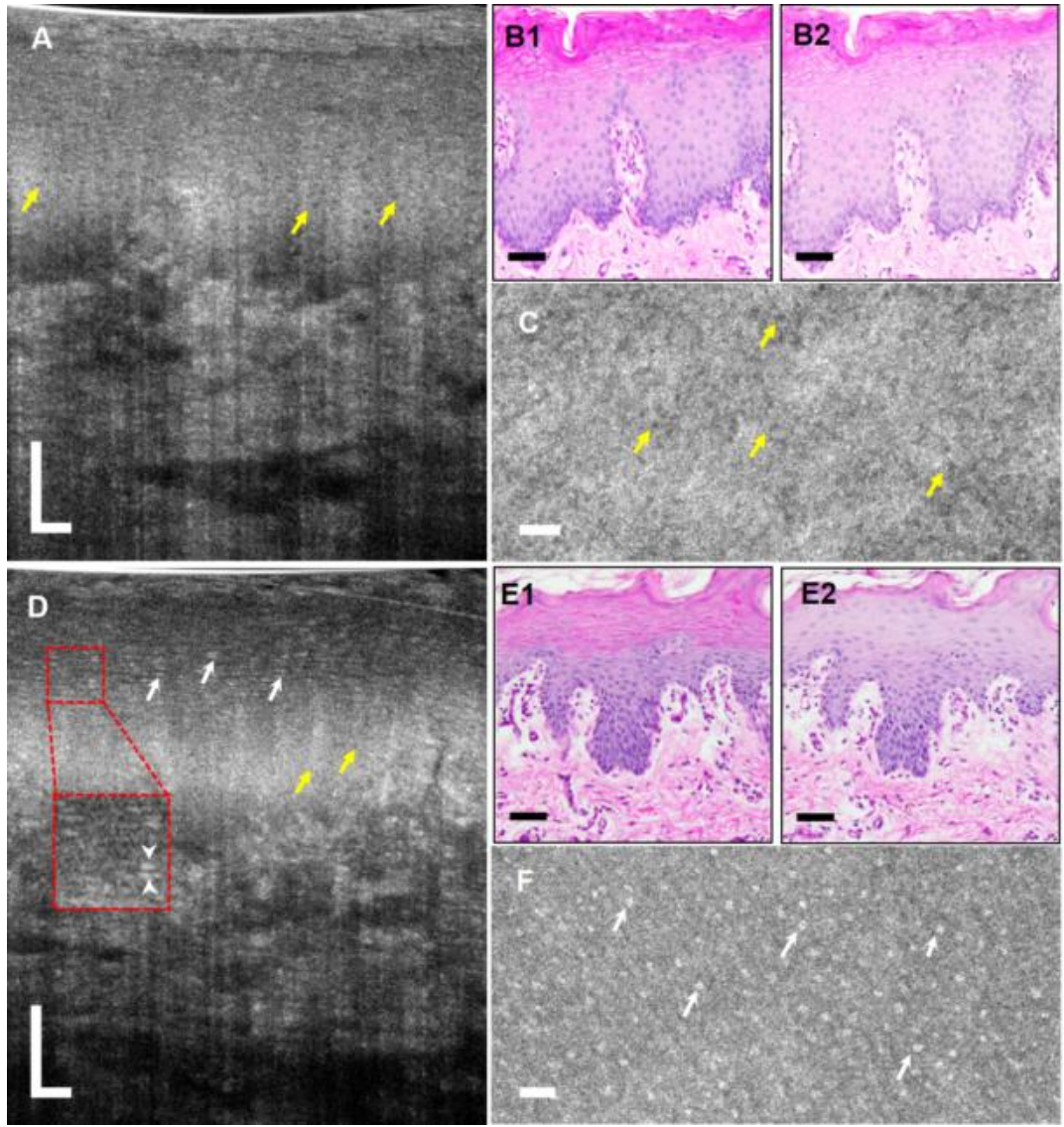

**Figure 4.** μOCT images of swine floor-of-mouth acquired *ex vivo* (Images with isotropically sized pixels could be found in **Figure S2**). (A) A representative cross-sectional μOCT image showing few high-scattering signals at the nucleocytoplasmic boundary (see **Video S1**, a); (B1) the corresponding histology with PAS staining and (B2) that with PAS-Diastase staining indicating cells with glycogen-negative cytoplasm and roundish nuclei; (C) *en face* μOCT view of the upper third of the epithelium. (D) A representative cross-sectional μOCT image presenting abundant high-scattering signals at the nucleocytoplasmic boundary (see **Video S1**, b); (E1) the corresponding histology with PAS staining and (E2) that with PAS-Diastase staining indicating cells with glycogen-positive cytoplasm and flattened nuclei; (F) *en face* μOCT view of the upper third of the epithelium. Yellow arrows represents low-scattering nuclei; white arrows and arrowheads indicate high-scattering nucleocytoplasmic boundary. The contrast of all images was adjusted with imageJ for the optimal visualization of nucleus and nucleocytoplasmic boundary. Scale bars, 50 μm.

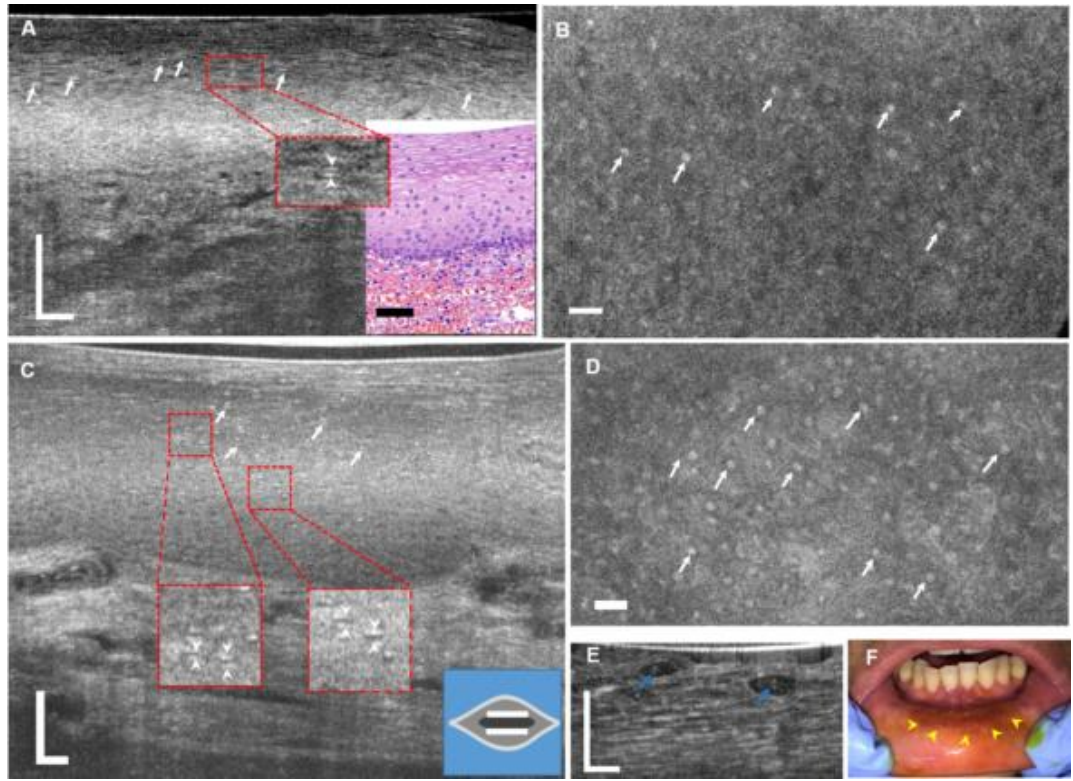

**Figure 5.** µOCT images of high-scattering signals at the nucleocytoplasmic boundary in the human esophageal and labial mucosal epithelium (Images with isotropically sized pixels could be found in **Figure S3**). (A) µOCT cross-sectional and (B) *en face* images of the human oesophagus acquired *ex vivo*; lower-right inset in (A) is the corresponding histology. (C) µOCT cross-sectional and (D) *en face* images of the human labial mucosa obtained *in vivo* showing high-scattering nucleocytoplasmic boundaries (see **Video S2&S3**); lower-right inset in (C) is a schematic of scattering profile of squamous cells. Red dashed boxes in (A&C) are 2× zoom view of selected areas; all white arrows and arrowheads indicate the high-scattering nucleocytoplasmic boundary. (E) µOCT image of the most superficial layers of human labial mucosal epithelium *in vivo* and the blue arrows suggest pyknotic nuclei or cell debris encircled by bright cell borders. (F) A photograph of human labial mucosa with Lugol's staining marked by yellow arrowheads. The contrast of all images was adjusted with imageJ for the optimal visualization of nucleus and nucleocytoplasmic boundary. Scale bars, 50 µm.

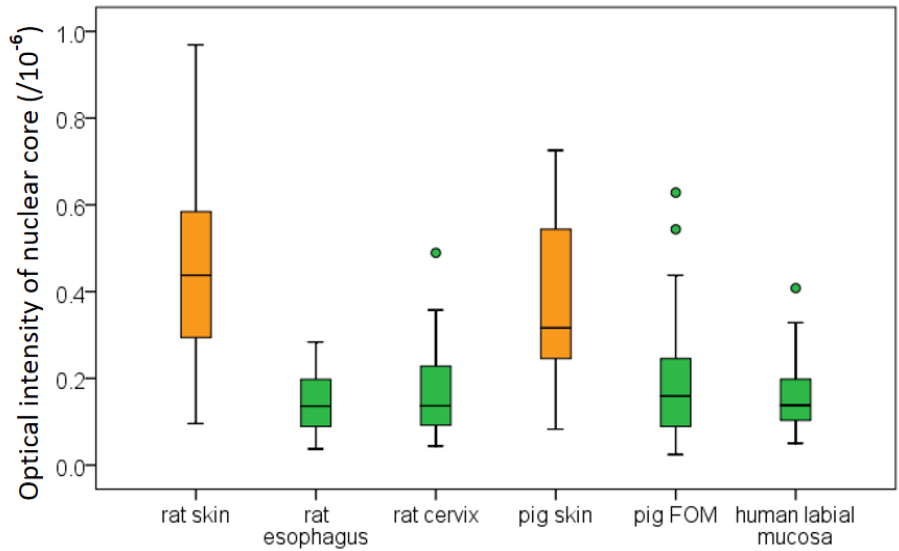

**Figure 6.** A boxplot displays the optical intensity of nuclear cores among mammalian SSE.

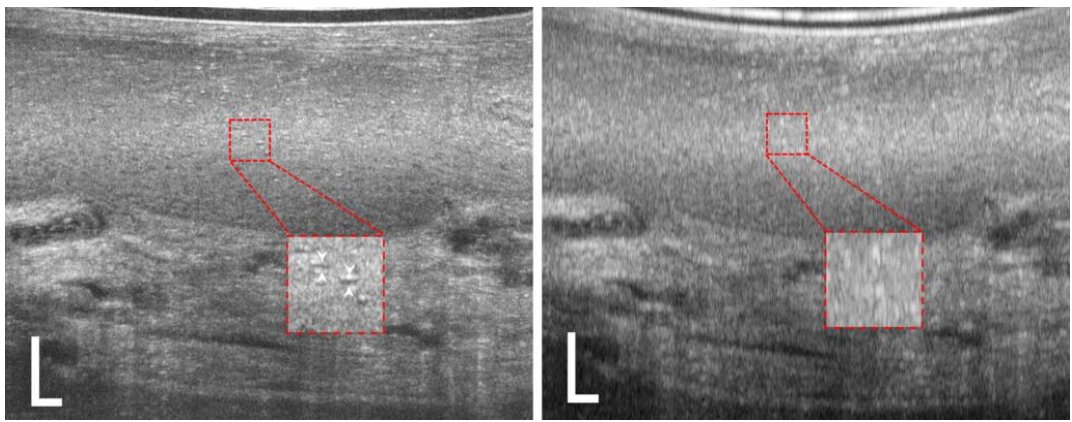

**Figure 7.** Cross-sectional µOCT images of human lalial mucosa. (A) Low-scattering nuclear cores sandwished by the high-scattering nucleocytoplasmic boundary are readily detectable by µOCT with an axial resolution of 1.28 µm in tissue; (B) The low-scattering nuclear cores can not be recognized leaving a vague bright dot of the nuclear region when the axial resolution is algorithmically degraded to 3.7 µm in tissue. Scale bars, 50 µm.

**Table 1.** Optical intensity of the nuclear core and cytoplasm in different mammalian SSE[†]

| SSE Type | nuclear core (mean ± SD) × $10^{-6}$ | cytoplasm (mean ± SD) × $10^{-6}$ | $P$[§] |
|---|---|---|---|
| Rat skin | 0.45 ± 0.22 | 9.10 ± 5.83 | < 0.001[*] |
| Rat esophagus | 0.14 ± 0.06 | 1.11 ± 0.30 | < 0.001[*] |
| Rat cervix | 0.16 ± 0.10 | 2.23 ± 1.63 | < 0.001[*] |
| Swine skin | 0.38 ± 0.17 | 5.19 ± 1.91 | < 0.001[*] |
| Swine floor-of-mouth (PAS -) | 0.19 ± 0.13 | 1.54 ± 0.61 | < 0.001[*] |
| Human labial mucosa | 0.15 ± 0.07 | 1.38 ± 0.49 | < 0.001[*] |

[†]Swine floor-of-month (PAS +) and human esophagus were not included due to insufficient data;
[§]Independent-samples t-test;
[*]The mean difference is statistically significant at the 0.05 level.

**Graphical Abstract**

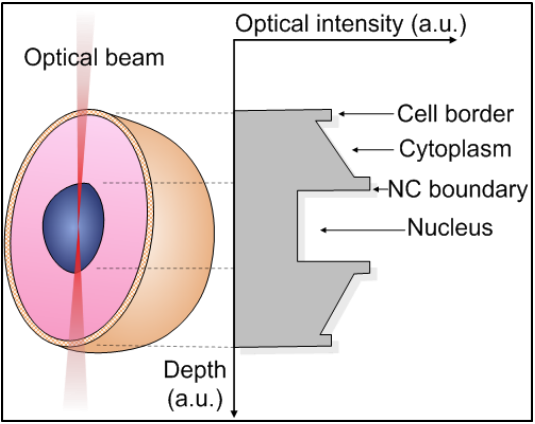

**Supporting Information**

**Supplementary figures and figure captions**

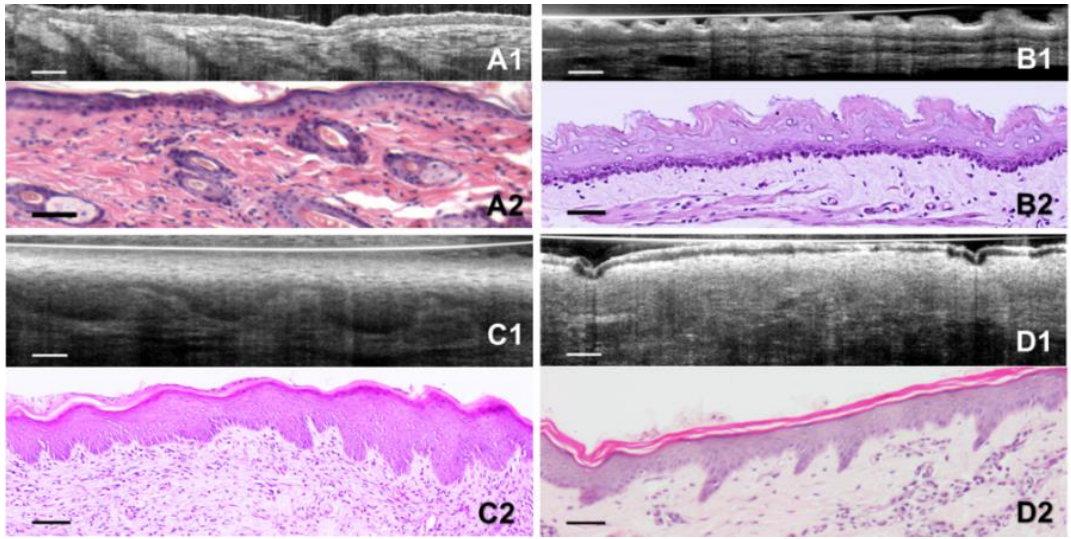

**Figure S1.** μOCT images and corresponding histology of the orthokeratinized stratified squamous epithelium in rat and swine acquired *ex vivo*. (A1-A2) rat skin, (B1,B2) rat esophagus, (C1,C2) rat cervix, (D1,D2) skin of white Cross-Landrace swine. Scale bars, 50 μm.

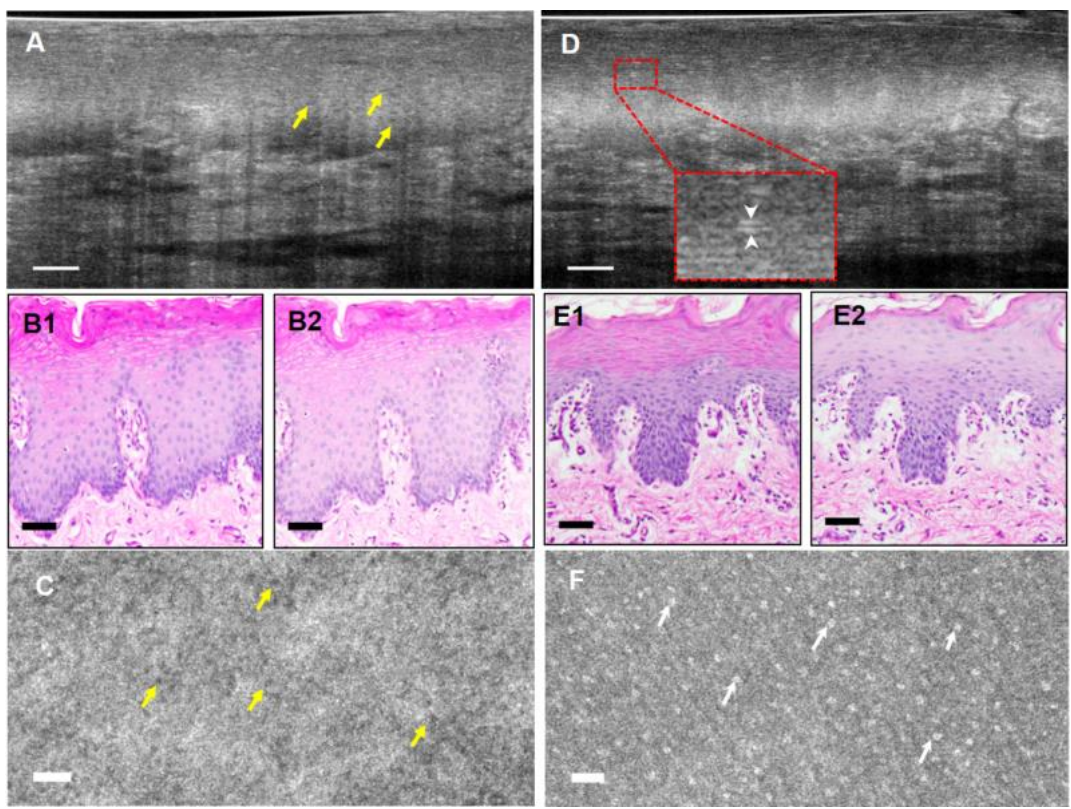

**Figure S2**. µOCT images of swine floor-of-mouth acquired *ex vivo*. (A) A representative cross-sectional µOCT image showing few high-scattering signals at the nucleocytoplasmic boundary; (B1) the corresponding histology with PAS staining and (B2) that with PAS-Diastase staining; (C) *en face* µOCT view of the upper third of the epithelium. (D) A representative cross-sectional µOCT image presenting abundant high-scattering signals at the nucleocytoplasmic boundary; (E1) the corresponding histology with PAS staining and (E2) that with PAS-Diastase staining; (F) *en face* µOCT view of the upper third of the epithelium. Red dashed box is 3× zoom view of selected areas. Yellow arrows represents low-scattering nuclei; white arrows and arrowheads indicate high-scattering nucleocytoplasmic boundary. Scale bars, 50 µm.

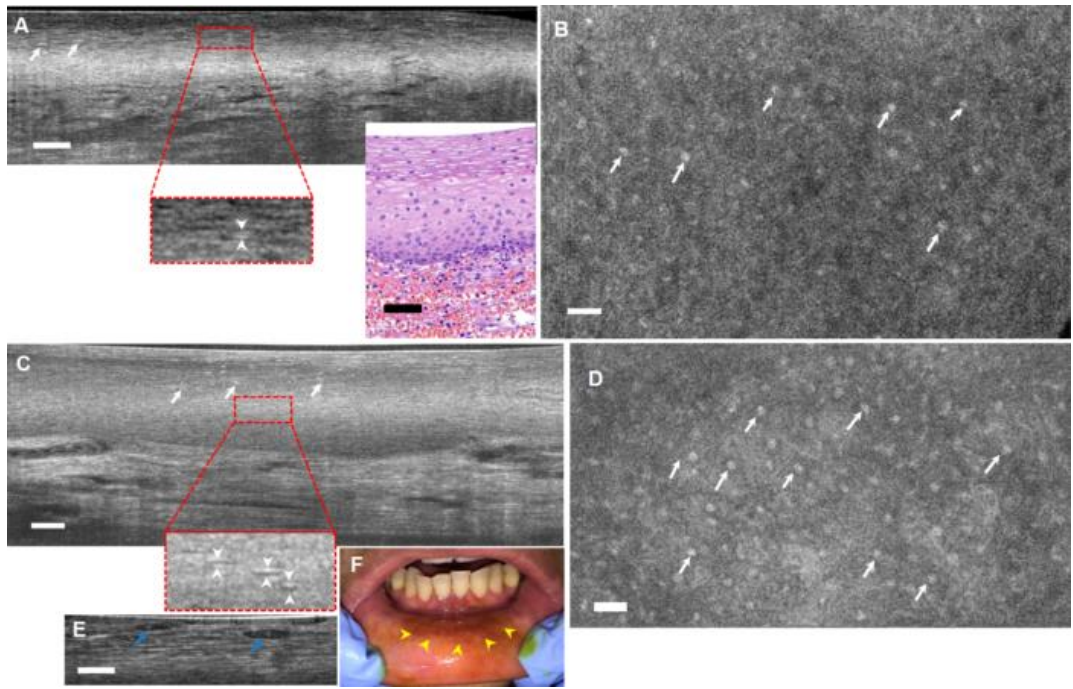

**Figure S3**. µOCT images of high-scattering signals at the nucleocytoplasmic boundary in the human esophageal and labial mucosal epithelium. (A) µOCT cross-sectional and (B) *en face* images of the human oesophagus acquired *ex vivo*; lower-right image is the corresponding histology. (C) µOCT cross-sectional and (D) *en face* images of the human labial mucosa obtained *in vivo* showing high-scattering nucleocytoplasmic boundaries. Red dashed boxes are 3× zoom view of selected areas; all white arrows and arrowheads indicate the high-scattering nucleocytoplasmic boundary. (E) µOCT image of the most superficial layers of human labial mucosal epithelium *in vivo* and the blue arrows suggest pyknotic nuclei or cell debris. (F) A photograph of human labial mucosa with Lugol's staining marked by yellow arrowheads. Scale bars, 50 µm.

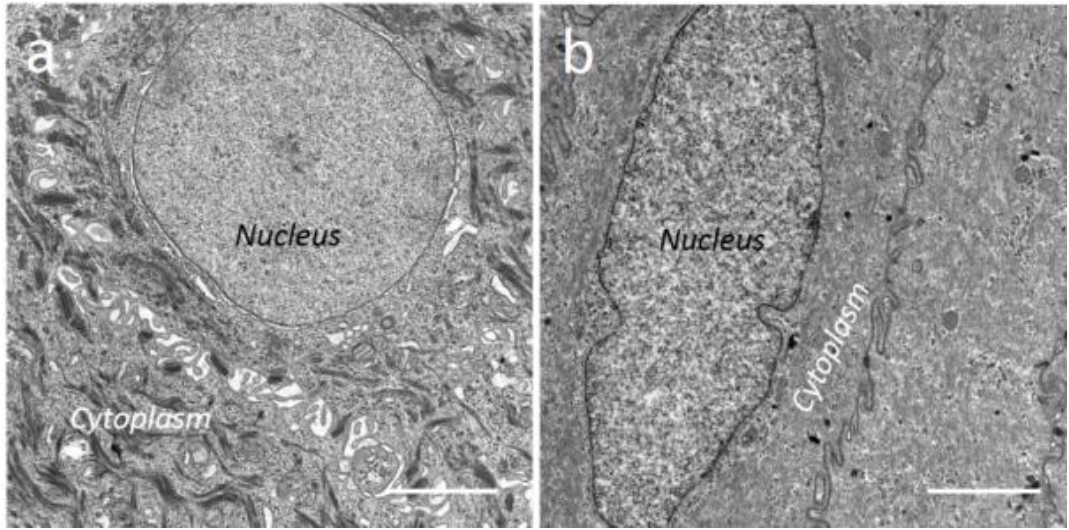

**Figure S4**. Representative TEMs of keratinocytes from swine keratinized skin (a) and nonkeratinized floor of month (b). Scale bars, 2 µm.

**Supplementray video captions**

**Video S1.** Cross-sectional µOCT images of a swine floor of month from a glycogen-poor (a) and glycogen-rich (b) area respectively acquired *ex vivo*. The glycogen-rich area (b) demonstrated more readily detectable high-scattering nucleocytoplasmic boundaries in the upper half epithelium. EP: epithelium; LP:lamina properia. Scale bars, 50 µm.

**Video S2.** Cross-sectional µOCT images of a human labial mucosa acquired *in vivo*. Paired, high-scattering nucleocytoplasmic boundary sandwiching a low-scattering nuclear core could be readily detected particularly in the upper half epitelial layers. EP: epithelium; LP:lamina properia. Scale bars, 50 µm.

**Video S3.** En face µOCT views of a human labial mucosa acquired *in vivo*. Abunant bright dots against relatively low-scattering cytoplasmic background could be recognized in the upper half epithelial layers. Scale bars, 50 µm.